\documentclass[a4,preprint]{iopart}
\usepackage{amsfonts}
\usepackage{latexsym}

\newtheorem{propo}{Proposition}
\newtheorem{teor}{Theorem}

\usepackage{iopams}

\begin{document}

\title[]{A simple spectral condition implying separability for states of bipartite quantum systems}
\author{ M E Gabach Cl\'ement and  G A Raggio}
\address{FaMAF, Universidad Nacional de C\'{o}rdoba, X5000 C\'{o}rdoba, Argentina}
 
\ead{raggio@famaf.unc.edu.ar}

\date{December 26, 2005}

\begin{abstract}
We give a simple spectral condition in terms of the ordered eigenvalues of the state of a bipartite quantum system which is sufficient for separability.
\end{abstract}

We consider quantum systems where the underlying Hilbert space ${\cal H}$ is the tensor product of two finite dimensional Hilbert spaces. A state of such a system is identified with a density operator, and is  said to be separable if it can be written as a convex sum of pure product states of the system; that is to say vector states where the vectors are product vectors. The separable states form a convex subset of the states of the system.\\
For the simplest bipartite composite system we have the following result
\begin{teor}  If the eigenvalues $\lambda_1 \geq \lambda_2\geq \lambda_3 \geq \lambda_4$ of the two qubit state $\rho$ satisfy $ 3 \lambda_1 +\sqrt{2}\lambda_2 + (3-\sqrt{2})\lambda_3 \leq 2$,
  then $\rho$ is separable.\end{teor} 
The states satisfying the inequality have spectra in the simplex spanned by the spectra (always written taking into account multiplicities and nonincreasingly) $(1/2,1/6,1/6,1/6)$, $((2+\sqrt{2})/8, (2+\sqrt{2})/8,(2-\sqrt{2})/8,(2-\sqrt{2})/8)$, $(1/3,1/3,1/3,0)$ and $(1/4,1/4,1/4,1/4)$.\\
The method used to prove the above result also gives a different proof of the following result given in \cite{Ra} (Theorem 3)
\begin{teor}If the eigenvalues $\lambda_1 \geq \lambda_2\geq \cdots  \geq \lambda_d$ of the  state $\rho$ of a bipartite quantum system of dimension $d$  satisfy $ 3 \lambda_d +(d-1)\lambda_{d-1} \geq 1$,
  then $\rho$ is separable.\end{teor} 
Both results provide simple spectral criteria ensuring separability. In the case of two qubits ($d=4$) Theorem 2 is much weaker than Theorem 1.\\

Before proceeding to the proofs, we compare theorem 1 with other available results of the same nature--that is: conditions on the spectrum implying separability of the state. Given a state $\rho$ of a $d$-dimensional bipartite quantum system, we let $spec(\rho )= ( \rho_1, \rho_2,\cdots , \rho_d )$ denote the vector of repeated eigenvalues of $\rho$ enumerated so that $\rho_1\geq  \rho_2\geq \cdots \geq  \rho_d$. In the two qubit case, let $\Sigma = \{ ( \lambda_1, \lambda_2, \lambda_3, \lambda_4 ):\; \lambda_1\geq \lambda_2\geq\lambda_3\geq\lambda_4\geq 0\;,\; \lambda_1 +
\lambda_2+ \lambda_3+\lambda_4=1\}$ be all possible state spectra. Theorem 1 asserts that if $spec (\rho )$ lies in $ {\cal A}:=\{\lambda \in \Sigma :\; 3\lambda_1+\sqrt{2}\lambda_2+(3-\sqrt{2})\lambda_3\leq 2 \}$, then   $\rho $ is separable.  One of the first and most useful results of this nature is that of \cite{Zy}: If $tr ( \rho^2) \leq 1/3$ then $\rho$ is separable. In terms of spectra this is: $spec ( \rho ) \in {\cal B}:=\{ \lambda \in \Sigma :\; \lambda_1^2+\lambda_2^2+\lambda_3^2+\lambda_4^2\leq 1/3\}$ implies $\rho$ is separable. Although ${\cal A}\cap {\cal B}$ is quite large, ${\cal A}\neq {\cal B}$, and the conditions defining ${\cal A}$ and ${\cal B}$ capture different (convex) sets of separable states. To see this, observe that $\lambda\equiv ((2+\sqrt{2})/8, (2+\sqrt{2})/8,(2-\sqrt{2})/8,(2-\sqrt{2})/8)\in {\cal A}$ but $\lambda_1^2+\lambda_2^2+\lambda_3^2+\lambda_4^2=3/8 > 1/3$. Moreover, $ (\sqrt{2}/3)(3/4,1/4,0,0) + (1-(\sqrt{2}/3))(1/4,1/4,1/4,1/4) $ lies in ${\cal B}$ but not in ${\cal A}$.
The determination of ``maximally entangled'' states of two qubits by Verstraete, Audenaert and De Moor, \cite{Ve}, has the following as a consequence\footnote{We thank H. Vogts and K.
\.{Z}yczkowski for bringing  \cite{Ve} to our attention.}: let ${\cal C}:= \{ \lambda \in \Sigma:\; \lambda_1-\lambda_3 - 2 \sqrt{\lambda_2 \lambda_4} \leq 0 \}$; then $\rho$ is separable  if $spec ( \rho )\in {\cal C}$. The inequality $\sqrt{(ta+(1-t)b)(tc+(1-t)d)} \geq t \sqrt{ac}+(1-t)\sqrt{bd}$ valid for $0 \leq t\leq 1$ and $a,b,c,d \geq 0$ shows inmediately that ${\cal C}$ is convex. By the results of \cite{Ve}, one has ${\cal B}\subset {\cal C}$\footnote{A direct proof goes as follows. It suffices to show that if $\lambda \in {\cal B}$ with $\lambda_1^2+  \lambda_2^2+\lambda_3^2+\lambda_4^2=1/3$ then $\lambda \in {\cal C}$. Now putting $\lambda_4=1-\lambda_1-\lambda_2-\lambda_3$ in the previous identity, we obtain
\[ \lambda_1^2+\lambda_2^2+ \lambda_3^2-(\lambda_1+\lambda_2+\lambda_3)+(\lambda_1\lambda_2+\lambda_1\lambda_3+\lambda_2\lambda_3)=-1/3\;.\]
 Then, $(\lambda_1-\lambda_3)^2-4 \lambda_2\lambda_4=\lambda_1^2+\lambda_3^2 -6\lambda_1\lambda_3-4\lambda_2+4 \lambda_2^2+4(\lambda_1\lambda_2+\lambda_1\lambda_3+\lambda_2\lambda_3)$. Using the displayed identity to eliminate the
summand $4(\lambda_1\lambda_2+\lambda_1\lambda_3+\lambda_2\lambda_3)$, we obtain $(\lambda_1-\lambda_3)^2-4 \lambda_2\lambda_4= -3(\lambda_1+\lambda_3-2/3)^2 \leq 0$, and this proves that $\lambda \in {\cal C}$.} One verifies that the four vertices of ${\cal A}$ given after the statement of theorem 1 lie in ${\cal C}$ so that ${\cal A} \subset {\cal C}$ because ${\cal A}$ is the convex hull of its four vertices. \\

The proof of the two stated results uses certain tools developped in \cite{Ra} which we  briefly present. Consider the maximally mixed state $\tau = {\bf 1}/ dim ( {\cal H})$, then $\tau$ factorizes  over the two factors of ${\cal H}$ so that $\tau$ is a separable state. Consider the segment with endpoints $\rho$ and $\tau$: $ \rho_t = t \cdot \rho +(1-t )\cdot \tau $, $0 \leq t \leq 1$.
The modulus of separability $\ell$ \cite{Ra}, measures how far  you can go towards $\rho$ beginning at $\tau$ until you lose separability: $ \ell ( \rho ) = \sup \{ t : \; \rho_t \mbox{ is separable}\}$.
The quantity $(1/\ell ) -1$ was studied by Vidal and Tarrach \cite{ViTa} as the ``random  robustness of entanglement''. It can be shown (\cite{ViTa, Ra}) that
the supremum is a maximum; that $\rho_t$ is separable iff $t \leq \ell ( \rho )$; that $\ell ( \rho ) >0$; and that $1/\ell$ is a convex map on the states: for states $\rho , \phi$ and $0\leq s \leq 1$,
\begin{equation}\ell ( s.\rho +(1-s)\cdot \phi   ) \geq \left(  \frac{s}{\ell ( \rho )} + \frac{1-s}{\ell ( \phi )} \right)^{-1} \label{conv}\;.\end{equation}
 The other ingredient is the so called  gap-representation of a state introduced in \cite{Ra}.
Let $spec ( \rho ) = ( \lambda_1, \lambda_2,\cdots , \lambda_d )$ and $\rho = \sum_{j=1}^d \lambda_j \cdot \rho_j$ be a spectral decomposition of $\rho$ where the $\rho_j$ are  pairwise orthogonal pure vector states. Define $ \mu_j =j(\lambda_j-\lambda_{j+1})$, $j=1,2,\cdots, d-1$; $ \widehat{\rho}_j = j^{-1}\sum_{m=1}^j \rho_m$, $j=1,2,\cdots , d$.
Notice that $\sum_{j=1}^{d-1} \mu_j = 1-d \lambda_d$; that $\widehat{\rho}_d = \tau$; and that
\[ spec ( \widehat{\rho}_j) =  (\underbrace{1/j,1/j, \cdots , 1/j}_{j},0,\cdots , 0 ) \equiv e^{(j)}\;.\]
So $\widehat{\rho}_1$ is pure.  Then a gap-representation of $\rho$ is $ \rho = \sum_{j=1}^{d-1}\mu_j \cdot \widehat{\rho}_j + d\ \lambda_d\cdot \tau $.
Noticing that $\lambda_d =1/d$ iff $\rho=\tau$, we assume that this is not the case and write
\[  \rho = (1-d\lambda_d)\cdot \omega + d \lambda_d \cdot \tau \;,\;\;\; \omega =  \sum_{j=1}^{d-1}\frac{\mu_j}{1-d\lambda_d}\cdot \widehat{\rho}_j \;.\]
By the results mentioned,  $\rho$ is separable iff
 \begin{equation} (1-d\lambda_d) \leq \ell ( \omega ) \;.\label{crit} \end{equation}   
 Applying (\ref{conv}) to the state $\omega $ in its gap representation, we have
\[ \ell ( \omega ) \geq \left( \sum_{j=1}^{d-1}\frac{\mu_j}{(1-d\lambda_d)\ell ( \widehat{\rho}_j)} \right)^{-1} = (1-d\lambda_d) \left( \sum_{j=1}^{d-1}\frac{\mu_j}{\ell ( \widehat{\rho}_j)} \right)^{-1} \;;\]
thus (\ref{crit}) is satisfied (and thus $\rho$ is separable) if $\sum_{j=1}^{d-1} \mu_j / \ell ( \widehat{\rho}_j) \leq 1$.
We can replace the $\ell ( \widehat{\rho}_j)$ by lower bounds.
\begin{propo}If $\ell ( \widehat{\rho}_j)\geq p_j\geq 0$ for $j=1,2,\cdots, d-1$ and $\sum_{j=1}^{d-1} \mu_j/p_j \leq 1$ then $\rho$ is separable.
\end{propo}

The prime reason for introducing the gap-representation is that not only the last summand $\tau$  but also the second last $\widehat{\rho}_{d-1}$ are separable. This follows from a result of Gurvits and Barnum \cite{GuBa}: If $tr ( \phi^2) \leq 1/(d-1)$ for a bipartite composite system of dimension $d$ then $\phi$ is separable. Now indeed $tr ( \widehat{\rho}_{d-1}^2)=1/(d-1)$.\\
The least possible modulus of separability has been  computed by Vidal and Tarrach \cite{ViTa}: $\inf \{ \ell ( \phi ) : \mbox{ $\phi$ a state }\} = 2/(2+d)$; the infimum is assumed at a pure state.
To prove theorem 2, put $p_1=p_2=\cdots =p_{d-2}=2/(2+d)$ and $p_{d-1}=1$ in the proposition.\\
Turning to theorem 1, consider the numbers $ \widehat{\ell}_j := \inf \{\ell (\phi): \; spec( \phi ) = e^{(j)}\}$,
which give the minimal moduli of separability for the states spanning all possible gap-representations.
Replacing $p_j$ by $\widehat{l}_j$ in the Proposition gives us a general inequality providing a sufficient condition for separability. No general information is available for the $\widehat{l}_j$ except the calculation of \cite{Ga} for two qubits where $\widehat{l}_1=1/3$, $\widehat{l}_2=1/\sqrt{2}$, and $\widehat{l}_3=1$. From this and the proposition one gets theorem 1. Since $\widehat{l}_1=1/3$, and $\widehat{l}_3=1$ follow from the results quoted above, we only give the calculation of $\widehat{l}_2$ in the appendix.

\appendix
\section{Calculation of $\ell$ for a two qubit state with spec=(1/2,1/2,0,0)}

\cite{Ga} gives a direct calculation of $\widehat{\ell}_1$, $\widehat{\ell}_2$ and $\widehat{\ell}_3$ using the Wootters Criterion \cite{Wo}. Recall that if $\rho$ is a state of a two qubit system, the Wootters operator $W$ asssociated to it is
$ W = (  \sqrt{\rho}(\sigma_y \otimes \sigma_y)\overline{\rho} (\sigma_y \otimes \sigma_y)\sqrt{\rho})^{1/2}$.
Here all operators are taken as matrices with respect to a product orthonormal basis. 
\[ \sigma_y = \left( \begin{array}{cc}
0&-i\\
i&0\end{array}\right) \;\]
and $\overline{\rho}$ is the complex conjugate of $\rho$ taken with respect to the basis which is  real.
 The Wootters Criterion is: $\rho $ is separable if and only if the (repeated) eigenvalues $w_1\geq w_2\geq w_3\geq w_4 $ of $W$ satisfy $w_1\leq w_2+w_3+w_4$.\\
We will calculate the modulus of separability for any state $\rho$ for which  $spec (\rho ) =(1/2,1/2,0,0)$ by calculating the spectrum of the Wootters operator associated to
$ \rho_t = t \rho +(1-t)\tau $, $0 \leq t \leq 1$.
The spectrum of $\rho_t$ consists of two double eigenvalues
$\alpha = (1+t)/4$ and $\beta = (1-t)/4$ (which coincide for $t=0$ where $\rho_0=\tau$). In order not to overload the notation we consider a density  operator $A$ with
$spec (A)= (\alpha, \alpha, \beta , \beta )$ where $\alpha +\beta =1/2$, and $\alpha \geq \beta \geq 0$; thus $1/4 \leq \alpha \leq 1/2$. The spectral decomposition of $A$ reads $ A = \alpha P+\beta P^{\perp}$,
where $P$ is an orthoprojection of rank $2$ and $P^{\perp}={\bf 1}-P$ is its orthocomplement, another orthoprojection of rank $2$. It follows that $ (\sigma_y \otimes \sigma_y)\overline{A} (\sigma_y \otimes \sigma_y) =
\alpha Q + \beta Q^{\perp}$,
 where $Q= (\sigma_y \otimes \sigma_y)\overline{P} (\sigma_y \otimes \sigma_y)$ is an orthoprojection of rank $2$ and $Q^{\perp}={\bf 1}-Q$. Using this one obtains for the square of the Wootters operator  associated to $A$ the formula
$ W^2  = \beta^2 {\bf 1} +\beta (\alpha -\beta )(P+Q)+(\alpha -\beta )(\sqrt{\alpha \beta }-\beta ) (PQ+QP)
 + ( \alpha^2 -\beta^2 -2 \sqrt{\alpha \beta }( \alpha - \beta )) PQP$.
Now since $P$, $P^{\perp}$, $Q$ and $Q^{\perp}$ are orthoprojections of rank $2$ in a four dimensional Hilbert space, we have three mutually exclusive alternatives for the subspaces $U$ and $V$ spanned by $P$ and $Q$ respectively: (1) $U\cap V=\{0\}$ which happens when and only when $Q={\bf 1}-P$ which is equivalent to $tr(PQ)=0$; (2) $dim ( U \cap V )=2$ which happens when and only when $Q=P$ which is equivalent to $tr(PQ)=2$; and (3) $dim (U\cap V) =1$ which happens when and only when there are unit vectors $\psi$, $\phi$ and $\chi$ in the four dimensional Hilbert space which satisfy $\langle \psi, \phi\rangle=\langle \psi ,\chi \rangle =0$ and $\mid \langle \chi, \phi \rangle \mid < 1$ such that $P= \mid \psi \rangle \langle \psi \mid + \mid \phi \rangle \langle \phi \mid$ and $Q= \mid \psi \rangle \langle \psi \mid + \mid \chi \rangle \langle \chi \mid$. One has $tr(PQ)= 1+ \mid \langle \chi,\phi \rangle \mid^2 $. This alternative is equivalent to $tr (PQ) \in [1, 2)$.\\

The three alternatives are distinguished by the value of $tr (PQ)$. For convenience we introduce the following characteristic geometric parameter $\xi = tr (PQ) -1 $,
which will determine the modulus of separability completely. We now distinguish the three possibilities.\\
(1) which occurs iff $\xi =-1$. Here $PQ=0$ allows one to compute $W^2 = \alpha \beta {\bf 1}$. The Wootters Criterion is satisfied and the associated state is separable.\\
(2) which occurs iff $\xi =1$. Here $P=Q$ allows one to calculate directly $W =A$, and the Wootters Criterion is just $\alpha \leq 1/2$ so the associated state is separable.\\
(3) which occurs iff $0 \leq \xi < 1$. We may assume that
\[ \psi = \left( \begin{array}{c} 1\\ 0\\ 0\\ 0
\end{array} \right) \;, \; \; \phi = \left( \begin{array}{c} 0\\ 1\\ 0\\ 0
\end{array}\right)\;,\;\; \chi = \left( \begin{array}{c} 0\\ \sqrt{\xi}\\ \eta_1\\ \eta_2
\end{array}\right)\;,\;\eta = \left( \begin{array}{c} \eta_1\\ \eta_2 \end{array}\right) \;,\]
where $\eta \neq 0$ because $\parallel \eta \parallel^2 = \parallel \chi \parallel^2 - \xi =1-\xi >0$.
We now partition ${\mathbb C}^4 = {\mathbb C}\oplus {\mathbb C}\oplus {\mathbb C}^2$, and doing the necessary matrix multiplications we get, from our previous formula for $W^2$,
\[ W^2 = \left( \begin{array}{ccc}
\alpha^2 & 0 & \langle 0\mid\\
0&\alpha \beta +\alpha (\alpha - \beta ) \xi & (\alpha - \beta )\sqrt{\xi\alpha\beta } \, \langle \eta \mid \\
\mid 0 \rangle & (\alpha - \beta )\sqrt{\xi\alpha\beta } \,\mid \eta  \rangle & \beta^2 {\bf 1}_2 +\beta (\alpha -\beta )\mid \eta \rangle \langle \eta \mid 
\end{array}  \right) \;.\]
It is now clear that $\alpha^2$ is an eigenvalue of $W^2$. The eigenvalue condition for an eigenvalue $\zeta$ to the eigenvector $x \oplus \mu $ for the lower right $3 \times 3$ block on ${\mathbb C}\oplus {\mathbb C}^2$ is
\begin{equation}
( \zeta -\alpha\beta -\alpha (\alpha -\beta )\xi ) x = (\alpha -\beta ) \sqrt{\xi \alpha \beta }\langle \eta , \mu \rangle \label{ew1}\end{equation}
\begin{equation} (\zeta - \beta^2 ) \mu = (\alpha - \beta ) ( \sqrt{\xi\alpha\beta } \, x + \beta \langle \eta, \mu \rangle ) \, \eta  \label{ew2} \;. 
\end{equation}
Putting $x=0$ and taking as we may $\mu \neq 0$ orthogonal to $\eta$, (\ref{ew1}) is satisfied and (\ref{ew2}) reduces to $(\zeta -\beta^2) \mu = 0$, thus $\beta^2$ is an eigenvalue of $W^2$.  We are now left with the problem of finding eigenvectors orthogonal to those already found. They are of the form $x \oplus c \eta$ with $x,c \in {\mathbb C}$. Inserting such eigenvectors into (\ref{ew1}) and (\ref{ew2}), the discussion of the solutions is tedious but straightforward. One obtains the two missing eigenvalues of $W^2$ to be
$  \zeta_{\pm} ( \alpha , \xi ) = \frac{\alpha}{2}(1-2\alpha) +\frac{\xi}{8}(4 \alpha -1)^2
 \pm \frac{4 \alpha -1}{4} \sqrt{2 \xi \alpha(1-2 \alpha )+ \xi^2 (2 \alpha -1/2)^2 }$.
Having the four eigenvalues of $A$ we must decide which is the largest. We have $\alpha \geq \beta$ by assumption, and clearly $\zeta_+ ( \alpha , \xi ) \geq \zeta_- ( \alpha , \xi ) $. Moreover, $\xi \mapsto \zeta_+ ( \alpha , \xi )$ is increasing and $\zeta_+ (\alpha , 1) = \alpha^2 $. Thus, $\alpha$ is the largest eigenvalue of $W$ and the Wootters Criterion reads: $  \alpha \leq  \beta + \sqrt{\zeta_+ ( \alpha , \xi )} +
\sqrt{\zeta_- ( \alpha , \xi )}$.
Manipulation of this inequality shows that it is equivalent to $ \alpha \leq \left( 1 + (1/\sqrt{2-\xi})\right)/4$.

Recalling that $\alpha =(1+t)/4$, we arrive at: If the two qubit state $\rho$ has $spec ( \rho ) =(1/2,1/2,0,0)$ then
\[ \ell ( \rho ) = \left\{ \begin{array}{lcl}
1 & ,&\mbox{ if $Q={\bf 1}-P$}\\
\frac{1}{\sqrt{3- tr (PQ)}} & ,& \mbox{otherwise}
\end{array}\right.\]
where $P$ is the spectral orthoprojection to the eigenvalue $1/2$ and $Q= (\sigma_y \otimes \sigma_y)\overline{P} (\sigma_y \otimes \sigma_y)$. Since $tr(PQ)\in [1,2]$ when
$Q\neq {\bf 1}-P$, we obtain $\widehat{\ell}_2 = \inf \{ \ell ( \rho ):\; spec ( \rho ) = (1/2,1/2,0,0)\} = 1/\sqrt{2}$.

\end{document}